\documentclass[12pt]{emulateapj}
\usepackage[]{natbib}
\usepackage{graphicx, amsmath}
\def\etal{{et~al.\null}}

\begin{document}

\title{{\it HST} Emission Line Galaxies  at $z \sim 2$: The Mystery of Neon}
\author{Gregory R. Zeimann \altaffilmark{1,2}, Robin Ciardullo \altaffilmark{1,2}, Henry Gebhardt \altaffilmark{1,2}, Caryl Gronwall \altaffilmark{1,2}, Alex Hagen \altaffilmark{1,2}, Jonathan R. Trump \altaffilmark{1,2,3}, Joanna S. Bridge \altaffilmark{1,2}, Bin Luo \altaffilmark{1,2}, and Donald P. Schneider \altaffilmark{1,2}}
\altaffiltext{1}{Department of Astronomy \& Astrophysics, The Pennsylvania State University, University Park, PA 16802}
\altaffiltext{2}{Institute for Gravitation and the Cosmos, The Pennsylvania State University, University Park, PA 16802}
\altaffiltext{3}{Hubble Fellow}

\begin{abstract}
We use near-IR grism spectroscopy from the {\sl Hubble Space Telescope\/} to 
examine the strength of [Ne~III] $\lambda 3869$ relative to H$\beta$, [O~II] $\lambda 3727$
and [O~III] $\lambda 5007$ in 236 low mass ($7.5 \lesssim \log(M_*/M_{\odot}) \lesssim 10.5$)
star-forming galaxies in the redshift range $1.90 < z < 2.35$.  By stacking the data
by stellar mass, we show that the [Ne~III]/[O~II] ratios of the $z \sim 2$ universe are marginally
higher than those seen in a comparable set of local SDSS galaxies, and that 
[Ne~III]/[O~III] is enhanced by $\sim 0.2$~dex.  We consider the possible explanations for
this $\sim 4 \, \sigma$ result, including higher oxygen depletion out of the gas-phase, denser 
H~II regions, higher production of ${}^{22}$Ne via Wolf-Rayet stars, and the existence of a larger population of X-ray obscured AGN at $z \sim 2$ compared to 
$z \sim 0$.  None of these simple scenarios, alone, are favored to explain the observed line ratios.  We
conclude by suggesting several avenues of future observations to further explore the mystery of enhanced [Ne~III] emission.
\end{abstract}
\label{sec:abs}

\section{Introduction}
\label{sec:intro}
The coupling of multi-object, near-infrared spectrographs to ground-based 8-m class telescopes 
 and the {\sl Hubble Space Telescope} has
revolutionized our ability to probe the conditions of star formation in the high-redshift
universe.  The reason for this is simple:  at $z \sim 2$, all the key diagnostic emission
lines for density, temperature, and ionization state are shifted into the near-IR, where
atmospheric opacity and OH emission severely restrict our ability to measure faint lines.  
Consequently, virtually all the information we have about the physics of high-$z$ star formation
is based upon measurements of the bright lines of hydrogen (H$\alpha$ and H$\beta$),
oxygen ([O~II] $\lambda 3727$ and [O~III] $\lambda 5007$), nitrogen ([N~II] $\lambda 6584$),
and sulfur ([S~II] $\lambda\lambda 6717, 6731$).  
Galaxies at $z \sim 2$ have lower oxygen
abundances \citep[e.g.,][]{erb+06},  higher ionization parameters \citep[e.g.,][]{nakajima+14}, and harder ionization fields \citep{steidel+14} than $z \sim 0$ galaxies with similar stellar mass.
Moreover, by using a sample of 14 star-forming galaxies, \citet{shirazi+14} showed that, 
for a fixed stellar mass and specific star formation rate, the warm ionized gas in 
$z \sim 3$ star-forming regions is denser than at $z \sim 0.1$ by 
a full order of magnitude.

There is, however, one additional bright line that is rarely exploited.  [Ne~III]
$\lambda 3869$ is the isoelectronic equivalent of [O~III] $\lambda 5007$, with 
a strength that is often comparable to H$\beta$.   As both neon and oxygen are 
$\alpha$-process elements that are produced together in the same regions of 
Type~II supernovae, their abundance ratio should be fixed across population metallicity
\citep{woosley+95, timmes+95}, and indeed, that appears to be the case locally \citep{henry89,
izotov+06, izotov+11}.  However, Ne$^{++}$  has a slightly higher ionization potential than
O$^{++}$ (41.0 versus 35.1~eV), and its principal emission line at 3869~\AA\ has a slightly
higher excitation energy (3.2 versus 2.5~eV) and a much higher critical density for 
de-excitation ($\sim 10^7$ versus $10^6$~cm$^{-3}$) than its oxygen counterpart.   
The line therefore has significant diagnostic value, and can be used
to constrain the conditions of high-redshift star formation \citep{levesque+14}.

There have been few measurements of [Ne~III] $\lambda 3869$ in the high-redshift
universe.  Perhaps the most detailed is that by \citet{masters+14}, who used co-added
spectra to show that for $z = 1.3$  and $z = 2.3$ galaxies, [Ne~III] traces [O~III] with 
a strength that is similar but perhaps elevated ($\sim 2 \, \sigma$ offset) over that seen in 
the local universe.    The sample size of this program was small (23 galaxies), however, and the
diagnostic value of neon was not investigated.

Here we use near-IR spectroscopy from the {\sl Hubble Space Telescope (HST)\/} and a wealth 
of ancillary photometric data to examine the emission line ratios of [O~III]/H$\beta$, [Ne~III]/[O~II], 
and [Ne~III]/[O~III] in 236 star-forming galaxies in the redshift range $1.90 < z < 2.35$.  
In \S2, we discuss the data acquired over a $\sim 350$ arcmin$^2$ region of the 
GOODS-N, GOODS-S, and COSMOS fields, and the processing techniques needed to 
measure individual emission line ratios.  In \S3, we describe the procedures used to identify a
complete, [O~III]-selected sample of $z \sim 2$ galaxies, place our sample in the context of
other $z \sim 2$ surveys, and identify local systems with similar
stellar masses and star formation rates.  In \S4, we present our measurements of
[Ne~III] $\lambda 3869$ with respect to [O~II] $\lambda 3727$, [O~III] $\lambda 5007$,
and H$\beta$.  Since only $\sim 25\%$ of our sample have all four lines detected with a
signal-to-noise ratio greater than one, we also present measurements of a set of stacked spectra.  
In this section, we compare our
data to neon and oxygen measurements at $z \sim 0$, and demonstrate that the
average [Ne~III]/[O~III] ratio measured at $z \sim 2$ is significantly greater than that 
seen locally.  In \S 5, we discuss the possible sources of the high [Ne~III]/[O~III] ratio, including clumpy HII regions, 
hotter exciting stars, oxygen depletion out of the gas-phase, higher neon to oxygen abundances, and the existence
of Compton-thick
active galactic nuclei. 
We conclude by considering observations which would test these hypotheses. 

For this work we adopt a standard $\Lambda$CDM cosmology, with $\Omega_{M} = 0.3$,
$\Omega_{\Lambda} = 0.7$, and $H_0 = 70$~km~s$^{-1}$~Mpc$^{-1}$, and a \citet{kroupa01} initial mass function (IMF).  Unless stated otherwise, the equivalent widths (EW) discussed in this paper are quoted in the rest-frame.

\section{Data and Reductions}
\label{sec:obs}

Our study of emission line ratios in the $z \sim 2$ universe is focused on three 
$\sim 115$~arcmin$^2$ patches of sky in the COSMOS \citep{COSMOS}, GOODS-N, and GOODS-S \citep{GOODS} fields.  These regions have a wealth of photometric and
spectroscopic data available for analysis, including broadband photometry from {\sl HST\/}
and {\sl Spitzer,}  broad- and intermediate-band photometry from the ground, and
optical and near-IR slitless spectroscopy from {\sl HST.}  

\subsection{{\it HST} Spectroscopy}
\label{sec:spectroscopy}

Our rest-frame optical emission-line measurements come from the 3D-HST (GO-12177, 12328; P.I. P. van Dokkum) and AGHAST (GO-11600; P.I. B. Weiner) near-IR grism 
surveys with the WFC3 camera of the {\sl Hubble Space Telescope\/} 
\citep[][]{3DHST,weiner+14}.  The primary observations of these programs consisted of G141 slitless grism
spectroscopy covering the wavelength range $1.08~\mu{\rm m} < \lambda < 1.68~\mu$m at 
$R = 130$ resolution with a {$0\farcs 128$} spatial pixel scale.  Approximately 625~arcmin$^2$ was observed in the course of these surveys, 
including $\sim 80\%$ of the CANDELS footprint \citep{CANDELS}; 
when combined with accompanying direct images through the F140W filter of WFC3, these 
data provide full coverage of the rest-frame wavelengths 3700-5020~\AA\  for all 
$1.90 < z < 2.35$ galaxies with unobscured emission line fluxes brighter than 
$\sim 10^{-17}$~ergs cm$^{-2}$ s$^{-1}$.   This wavelength range includes the strong 
emission lines of [O~II] $\lambda 3727$, [O~III] $\lambda\lambda 4959,5007$, [Ne~III]
$\lambda\lambda 3869,3960$, and hydrogen (H$\beta$, H$\gamma$ and H$\delta$).  

An in-depth discussion of the reduction of these data can be found in \citet{zeimann+14}, 
but in brief, each grism observation was accompanied by a shallow ($\sim 200$~s) F140W exposure, which served to define the position of each object's wavelength zeropoint and 
spectral trace.   These images were combined using the standard procedures of 
{\tt MultiDrizzle} \citep{fruchter+09}, co-added with the deeper CANDELS F125W and F160W
frames, and processed by \textsc{SExtractor} \citep{bertin+96} to produce a master catalog of 
all objects brighter than $m_{AB} = 26$ and having more than five pixels above a $3 \, \sigma$ 
per pixel detection threshold.   Each object's corresponding grism spectrum was obtained
using the optimal extraction setting of version 2.3 of {\tt aXe} \citep{kummel+09}, as described in
the WFC3 Grism 
Cookbook\footnote[4]{http://www.stsci.edu/hst/wfc3/analysis/grism\_obs/cookbook.html} 
\citep[see][]{zeimann+14}.  At the conclusion of this process, the program 
{\tt aXe2web}\footnote[5]{http://axe.stsci.edu/axe/axe2web.html} was used to convert our final
list of objects into a webpage containing each object's 2-D and 1-D extracted spectra, its
$H$-band image, and a summary of its properties.  This provided an easy and efficient
way to maintain quality control and select subsamples of objects for science purposes.

\subsection{Optical/Near-IR Imaging}
\label{sec:imaging}
To measure the stellar masses and reddenings of our galaxies, we began with the
\textsc{SExtractor}-based photometric catalog of \citet{skelton+14}, which starts with 
deep, co-added F125W + F140W + F160W images from {\sl HST\/} and then adds in the 
results of 30 distinct ground- and space-based imaging programs.  The result is a 
homogenous, PSF-matched set of broad- and intermediate-band flux densities covering the 
wavelength range $0.35~\mu$m to $8.0~\mu$m over the entire region surveyed by the {\sl HST\/} grism\null.
In the COSMOS field, this dataset contains photometry in 44 separate bandpasses, with
measurements from {\sl HST, Spitzer,} Subaru, and a host of smaller ground-based telescopes.  
In GOODS-N, the data come from {\sl HST, Spitzer,} Keck, Subaru, and the Mayall telescope,
and include 22 different bandpasses, while in GOODS-S, six different telescopes
provide flux densities in 40 bandpasses.  For $z \sim 2$ systems, these data cover the 
rest-frame UV through the rest-frame near-IR and allow excellent estimates of stellar mass,
under the necessary assumptions about the underlying stellar population (i.e., 
stellar metallicity, star formation history, and attenuation law).

\section{Sample Selection and Measurements}
\label{sec:sample}

Our galaxy sample is nearly identical to that defined by \citet{zeimann+14}, who visually
inspected over 50,000 spectra in their program to identify $z \sim 2$ star forming
galaxies.  These galaxies satisfied three basic criteria:

$\bullet$  A secure redshift between $1.90 < z < 2.35$.  In this redshift interval, the emission
lines of [O~II] $\lambda 3727$, [Ne~III] $\lambda 3869$, H$\gamma$, H$\beta$, and
[O~III] $\lambda 4959, 5007$ all fall into the range of the G141 grism.  To be part of the
sample, a galaxy had to have at least two of these lines reliably detected.  At
the $R = 130$ resolution of the G141 grism, the [O~III] doublet is blended, though with
a distinctive, easily-identifiable shape (see Figure 1 of \citealp{zeimann+14}).   A total of 309 unique galaxies 
were identified through this condition.

$\bullet$ Minimal contamination from field sources.  Overlapping spectra are a significant
issue in slitless spectroscopy, as the dispersed order of one source frequently overlaps the
spectrum of another.  To estimate the importance of this effect, \citet{zeimann+14} used the
sizes and magnitudes of every object in the {\sl HST\/} grism \textsc{SExtractor} catalog to produce
a 2-D Gaussian model of every expected spectrum \citep{kummel+09}.  This procedure 
identified most of the spectral superpositions, but not all:  in particular, the algorithm
occasionally missed some regions where the systematics of contamination subtraction 
dominate the error in the continuum.  Consequently, an additional visual inspection
was needed to identify those objects where this systematic error was greater than the 
statistical error of the target spectrum.  This criterion eliminated 57 objects from the sample.

$\bullet$ A well-fit spectral energy distribution.  As described in \S 3.1, we obtained masses for
our $z \sim 2$ galaxies by fitting their spectral energy distribution to population synthesis
models via a Markov-Chain Monte Carlo code.  For 16 objects, this procedure did not
converge, due either to a misidentification of the asymmetric [O~III] doublet, or
poor broadband photometry.  These objects were excluded from our analysis, leaving
us with a final sample of 236 galaxies distributed over the GOODS-N, GOODS-S, and
COSMOS fields.

To understand the systematics of this sample, \citet{zeimann+14},  simulated a set of 
model emission-line spectra in the exact same manner as our program data.  These artificial
sources were randomly drawn from uniform distributions in redshift ($1.90 < z < 2.35$), log
metallicity  ($7 < 12 + \log({\rm O/H}) < 9$), and log H$\beta$ flux ($-18 < \log F_{{\rm H}\beta} 
< -16$~ergs s$^{-1}$~cm$^{-2}$), placed on mock grism frames, and ``observed'' in the 
exact same manner as the original data.  The results of this experiment can be found in 
Figure~2 of \citet{zeimann+14}.  In brief, for the GOODS-N and GOODS-S fields, our  
50\% and 80\% completeness limits for H$\beta$ flux measurements are $\sim 10^{-17}$~ergs~s$^{-1}$~cm$^{-2}$ and $\sim 3 \times 10^{-17}$~ergs s$^{-1}$~cm$^{-2}$, respectively, with little variation
with redshift or metallicity.   Due to its higher background, the COSMOS limits are shallower by 
a factor of $\sim 1.5$ \citep{3DHST}. 

Emission line fluxes were determined by fitting the continuum of each $z \sim 2$ spectrum with a 
fourth-order polynomial, while simultaneously fitting Gaussians of a common width 
to the emission lines of [O~II] $\lambda 3727$, [Ne~III] $\lambda 3869$, H$\gamma$, H$\beta$, 
and [O~III] $\lambda 4959,5007$.  Our choice of polynomial was
driven by the possible presence of a 4000~\AA\ break in the stellar continuum, though fits
using a series of first-order polynomials around each emission line yielded similar results. 
Also, we did not correct any of our fits for underlying Balmer absorption \citep{moustakas+06}.   
In the local universe, typical corrections for H$\beta$ are of the order of $\sim 4$~\AA\  equivalent
width (EW), but this number is a function of both population age and IMF \citep{groves+12}.  
As can be seen in Figure 3 of \citet{zeimann+14}, this factor is relatively small compared to 
the equivalent widths considered here.  Indeed, to verify that the effect is minor,
we repeated all our analyses while adding 4~\AA~EW to each of our H$\beta$
measurements.  This action has the effect of increasing all our H$\beta$ fluxes by an average of   
$\sim 10\%$, hence lowering our [O~III]/H$\beta$ ratios by the same amount.

 \subsection{Global Properties of the Sample}
 \label{sec:sample}
To place our neon measurements in context, we must first define the global properties of
our galaxy sample, particularly their star formation rates (SFRs) and stellar masses.  
The former parameter can be computed in two ways:  through the H$\beta$ line luminosity, which
measures the ionizing flux from $M > 15 M_{\odot}$ stars born in the past $\sim 10^7$~yr, and the
rest-frame UV continuum, which records stellar emission from $M > 5 M_{\odot}$ stars younger
than $\sim 10^8$~yr \citep{kennicutt+12}.  \citet{zeimann+14} investigated the behavior
of these two indicators in the same sample of galaxies considered here, and found that 
UV-based SFRs are less sensitive to shifts in galactic metallicity than Balmer emission.  We 
therefore adopted the \citet{zeimann+14} UV-based SFRs for our analysis.  These values are derived 
from the photometry of \citet{skelton+14}, and are based on the fits to the 1600~\AA\ flux density,
assuming a power-law slope to the UV continuum, the \citet{calzetti01} obscuration law, and the local SFR calibration \citep{hao+11, murphy+11, kennicutt+12}.

Our stellar masses are based upon the analysis of \citet{gebhardt+14}, who used the
Markov Chain Monte Carlo code \textsc{GalMC} \citep{acquaviva+11} to model the spectral energy distribution (SED)
of every galaxy in our sample.   In brief, the \citet{skelton+14} 
photometric measurements were fit to the 2007 version of population synthesis models of
\citet{bruzual+03}, using a \citet{kroupa01} IMF over the range 
$0.1 \, M_{\odot} < M < 100 \, M_{\odot}$, a \citet{calzetti01} obscuration law, and the 
prescription for emission lines given by \citet{acquaviva+11}, as updated by \citet{acquaviva+12}.  
Since stellar abundances are poorly constrained by broadband SED measurements, the metallicity 
of these models was fixed at $Z = 0.2 \, Z_{\odot}$, which is close to the median gas-phase
metallicity of our sample \citep{gebhardt+14}.   To avoid the emission from polycyclic aromatic hydrocarbons, 
all data points redward of rest-frame $3.3~\mu$m were excluded from the fits \citep{tielens}, as 
were measurements blueward of 1216~\AA, where the statistical correction for intervening 
Ly$\alpha$ absorption \citep{madau95} may not always be appropriate.   Finally, for simplicity, the
SFR of each galaxy was assumed to be constant with time. Although almost certainly not the 
case, this constraint does not strongly affect our estimates for the stellar mass.

\begin{figure}[htp] 
\includegraphics[width=.48\textwidth]{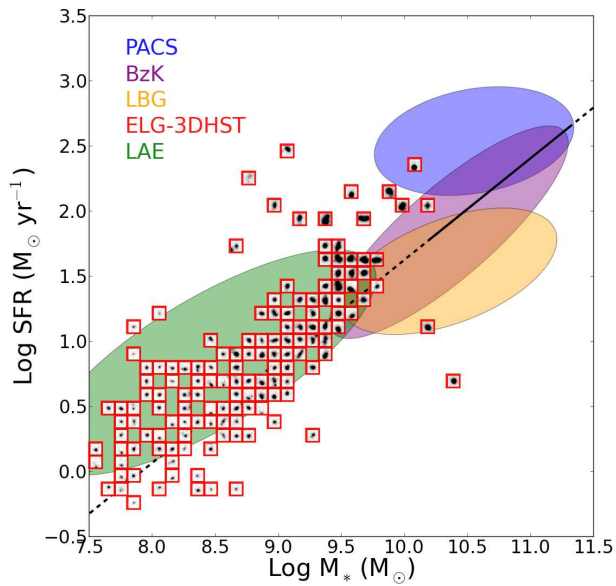}
\centering
\caption{A sample of our $z \sim 2$ star forming galaxies in the context of the epoch's ``galaxy 
zoo''.  The position of each F160W cut-out shows the system's stellar mass and star formation rate, 
as derived from its spectral energy distribution \citep{skelton+14}.   These ``normal''  emission-line
galaxies have lower stellar masses and star formation rates than $z \sim 2$ systems selected
via their {\sl Herschel\/} sub-mm emission \citep[blue PACS galaxies;][]{rodighiero+11},
Lyman break \citep[yellow galaxies;][]{erb+06}, or optical/IR colors
\citep[purple $BzK$ galaxies;][]{rodighiero+11}.    In this respect, they are similar to Ly$\alpha$ emitting 
galaxies \citep[green LAEs;][]{vargas+14, hagen+14}.  The solid black line is the star-forming ``main sequence'' of \citet{rodighiero+11}, 
while the dashed line extrapolates this relation.}
\label{fig:sfrmass}
\end{figure}

Figure~\ref{fig:sfrmass} compares the masses and SFRs of our galaxies to other
members of the $z \sim 2$ ``galaxy zoo''.  For the figure, literature SFRs and stellar masses
were converted from a \citet{salpeter55} IMF to a \citet{kroupa01} IMF by subtracting 
0.16 and 0.25 dex, respectively.  No correction was made from a \citet{chabrier03} IMF to a \citet{kroupa01} IMF as 
this correction is small ($\lesssim$6\%).   We highlight other $z \sim 2$ galaxy samples using ellipses defined 
to include $\sim90$\% of their galaxies.  In general, the galaxies
selected by the {\sl HST\/} grism are less massive and have lower star formation rates than
systems selected via their broadband continuum measurements.   In this respect, our emission-line
 galaxies are similar to Ly$\alpha$ emitting systems \citep[LAEs,][]{vargas+14, hagen+14}, and, 
like LAEs, the {\sl HST\/} emission-line sources tend to fall above the extrapolated SFR-stellar 
mass ``main sequence'' defined by \citet{rodighiero+11}.  

Another way to view the global properties of our sample of $z \sim 2$ grism galaxies
is through a comparison with local star-forming systems.   Such an analysis is not 
straightforward since, on average, $z \sim 2$ systems form stars ten times more rapidly 
than do local objects \citep[e.g.,][]{whitaker+14}.   Nevertheless, we can create an analogous
sample by applying luminosity, equivalent width, and stellar mass cuts to the 
Sloan Digital Sky Survey \citep[SDSS;][]{sdss-1} Data Release  7 \citep{sdss} MPA-JHU galaxy catalog\footnote[6]{www.mpa-garching.mpg.de/SDSS/DR7}.   Specifically, we selected objects with
extinction-corrected H$\beta$ luminosities $L_{{\rm H}\beta} \geq 3 \times 10^{40}$~ergs s$^{-1}$,
stellar masses between $7.5  < \log (M/ M_{\odot}) <  11.5$, and H$\beta$ equivalent widths 
greater than 5~\AA\ (so that the measurement of flux is not dominated by the correction for 
stellar absorption).  For inclusion in our comparison sample, we also required that the 
galaxies' emission line fluxes at [O~II] $\lambda 3727$, [Ne~III] $\lambda 3869$, H$\beta$, 
[O~III] $\lambda 5007$, and [O~III] $\lambda 4363$ all be at least twice as large as their flux 
error.  This latter criterion facilitated direct measurements of the systems' electron temperatures,
enabling a better understanding of their physical conditions \citep{AGN3}, while only reducing the 
sample by $\sim20$\%.

\begin{figure}[htp] 
\includegraphics[width=.48\textwidth]{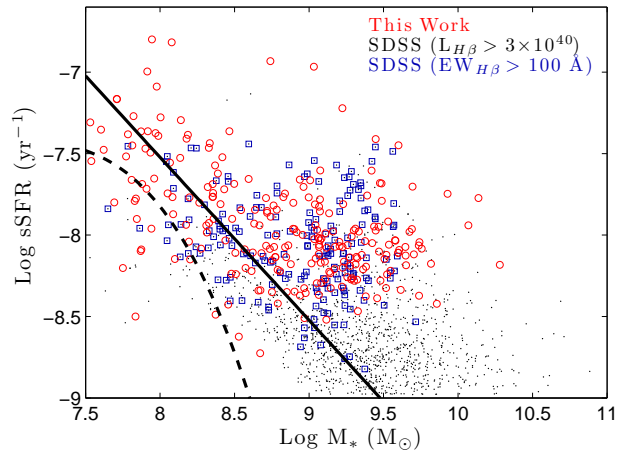}
\centering
\caption{The mass-specific star formation rate versus stellar mass diagram for our $z \sim 2$ grism galaxies (red open circles) and a local comparison sample from SDSS (black dots).   The blue open squares identify the subset of local galaxies with high H$\beta$ equivalent widths.  The black solid line illustrates 
our 50\% H$\beta$ flux limit converted into a rest-frame UV-based SFR at $z \sim 2.1$, using a
\citet{kroupa01} IMF and the $Z = 0.2 Z_{\odot}$ metallicity correction suggested by 
\citet{zeimann+14}.  The dashed black curve corresponds to our completeness limit
(F140W = 26) for a 2007 \citet{bruzual+03} stellar population model with $Z = 0.2 Z_{\odot}$ and 
a constant star formation history.  The closest local analogs to our emission-line
galaxies at $z \sim 2$ are the high-equivalent width objects studied by 
\citet{izotov+11}.}
\label{fig:ssfrmass}
\end{figure}

Our selection criteria resulted in the identification 2890 galaxies between 
$0.02 < z < 0.40$ with a median redshift of $z \sim 0.12$.  As illustrated in 
Figure~\ref{fig:ssfrmass}, our $z \sim 2$ grism galaxies have 
mass-specific star formation rates that are much higher than most of the comparison objects. 
Consequently, to obtain an even better set of analogs, we further restricted the local sample to
``young'' galaxies, i.e., systems with high H$\beta$ equivalent widths 
(EW$_{{\rm H}\beta} > 100$~\AA).   As Figure~\ref{fig:ssfrmass} demonstrates, this 
additional criterion creates a local sample that is quite similar to our $z \sim 2$ dataset, i.e., 
the distant star-forming galaxies identified by the {\sl HST\/} grism are 
comparable in both mass and specific star formation rate to the ``green 
pea'' galaxies \citep{cardamone+09} and the luminous blue compact galaxies identified by \citet{izotov+11} in their analysis of the SDSS spectral catalog.
By including lower equivalent width objects in our initial sample, however,
we did diversify the population: as the commonly used \citet{bpt81} diagnostic indicates, 
the larger comparison set contains both star forming regions and AGN, with some objects
being composites of the two populations.  This information will be useful for our subsequent
analysis.

\section{The Neon-Oxygen Line Ratios}
\label{sec:ratio}
\begin{figure*}[htp]
\vspace{-.4cm}
\includegraphics[width=.9\textwidth]{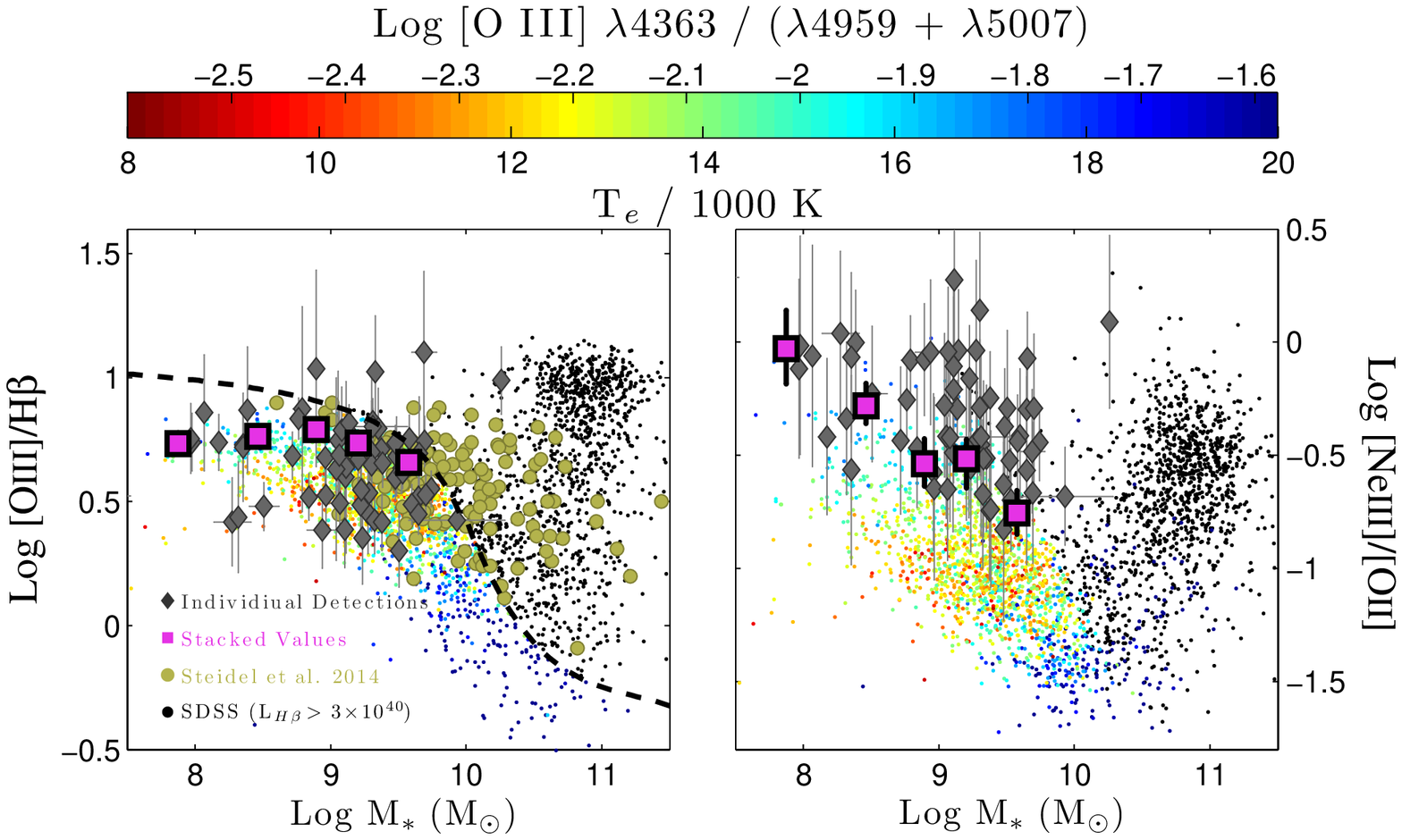}
\hspace{0cm}
\includegraphics[width=.9\textwidth]{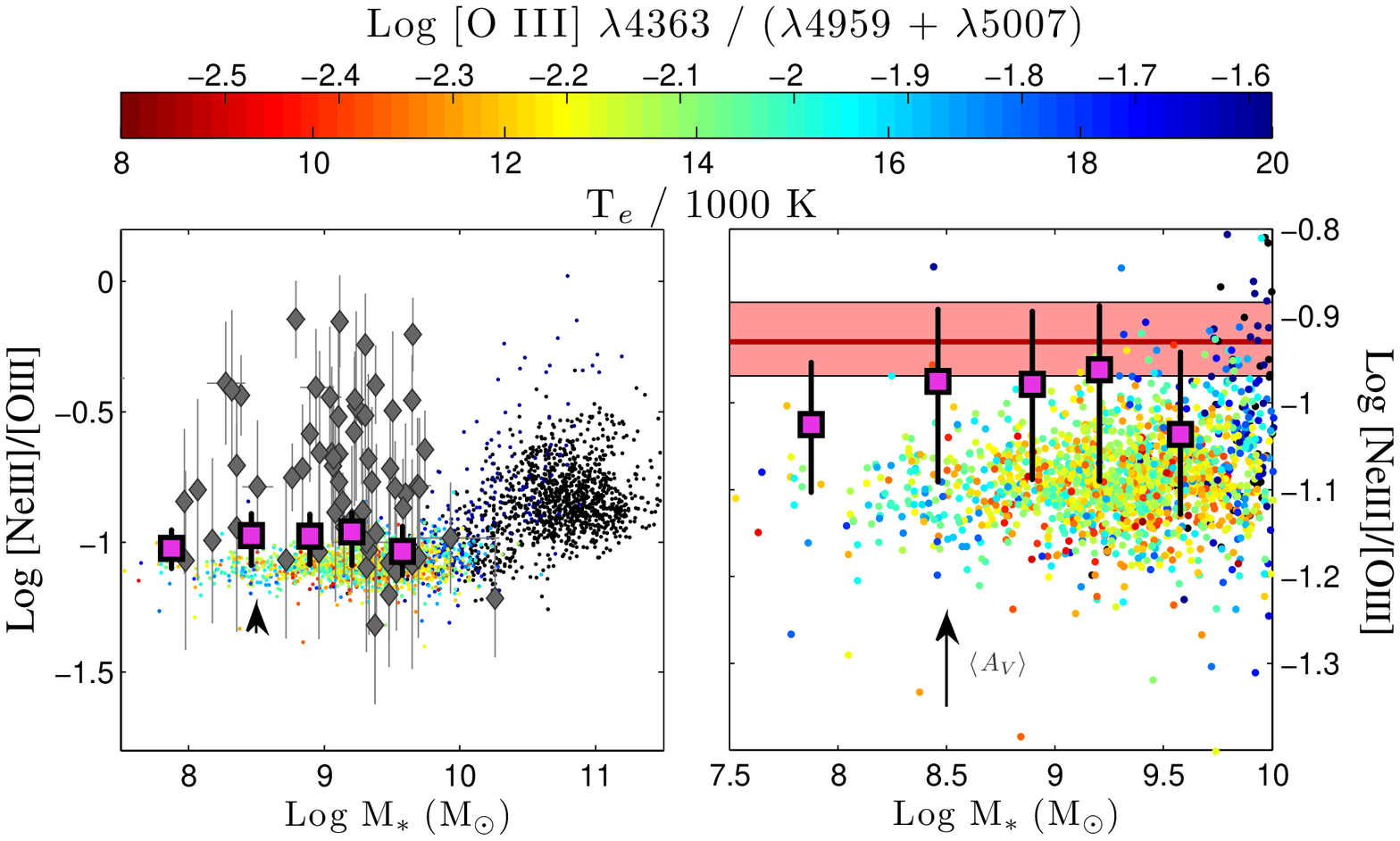}
\centering
\caption{\scriptsize{The observed ratios of [O~III]/H$\beta$ (top left), [Ne~III]/[O~II] (top right), and
[Ne~III]/[O~III] (bottom left) as a function of stellar mass, as derived from SEDs.   The small dots
show the more inclusive sample of local SDSS galaxies, color-coded by their [O~III]-based 
electron temperature, going from $T_e = 8000$~K (red) to 20000~K (blue), with black 
representing AGN-dominated systems \citep{juneau+14}.  The grey diamonds show the 
$z \sim 2$ {\sl HST\/}-grism galaxies with secure [Ne~III] detections (along with their $1 \, \sigma$
error bars), while the purple squares denote values obtained from our median stacks (which include non-detections).  The 
error bars for the stacks are the 16$^{\rm th}$ and 84$^{\rm th}$ percentiles of 500 bootstrap 
with replacement iterations.   In the [O~III]/H$\beta$ diagram, we have also added
measurements from the \citet{steidel+14} survey of more massive galaxies at $z \sim 2.3$ (gold dots), and have 
illustrated the dividing line between star-forming and AGN-dominated systems, as estimated from 
the MEx diagnostic for $z \sim 2$ and high-luminosity limits \citep{juneau+14}.   The arrow on the [Ne~III]/[O~III] diagram represents the
median extinction correction, based on the galaxies' UV slopes and a \citet{calzetti01}
obscuration law.  The bottom right panel is an expanded version of the [Ne~III]/[O~III] diagram,
where the individual $z \sim 2$ measurements have been replaced with their median value (red 
line), and their 16$^{\rm th}$ and 84$^{\rm th}$ percentiles (light red band). [Ne~III] $\lambda 3869$ is consistently brighter than expected from local samples.}}
\label{fig:ratiomass}
\end{figure*}

Figure~\ref{fig:ratiomass} plots the [O~III]/H$\beta$, [Ne~III]/[O~II], and [Ne~III]/[O~III] line ratios
against stellar mass for our sample of $z \sim 2$ star-forming galaxies, the \citet{steidel+14}
sample of massive (9 $\lesssim$ $\log$ M$_*$/M$_{\odot}$ $\lesssim$ 11.5) $z \sim 2.3$ star-forming (0.5 $\lesssim$ log SFR (M$_{\odot}$ yr$^{-1}$) $\lesssim$ 2) galaxies, and our more inclusive local 
comparison sample.   To illustrate the sequences, the local systems have been color-coded by
their [O~III]  $\lambda 4959 + 5007$ / [O~III] $\lambda 4363$ electron temperature \citep{AGN3}, 
and AGN-dominated systems, as determined by the mass-excitation diagnostic \citep{juneau+11,
juneau+14}, have been plotted in black.  From the figure, it is clear that,
although the measurement uncertainties are large, our $z \sim 2$ star-forming galaxies have
neon/oxygen line ratios that are significantly higher than that seen in the local universe.
In the [Ne~III]/[O~II] diagram, the {\sl HST\/} grism galaxies typically lie $\sim 0.2$~dex above
the sequence defined by local starbursting galaxies, while for [Ne~III]/[O~III], the median offset is
$\sim 0.3$~dex.  We note that, unlike the [O~III]/H$\beta$ and [Ne~III]/[O~II] line ratios, 
the [Ne~III]/[O~III] diagnostic is sensitive to reddening.  However, extinction corrections will only 
make the discrepancy worse by boosting [Ne~III] relative to [O~III];  if we convert the UV-based stellar reddenings
into nebular extinctions via the obscuration law of \citet{calzetti01}, the median offset 
increases to $\sim 0.4$~dex.  These high neon/oxygen ratios stand in striking contrast to the results
from [O~III]/H$\beta$, where the values for this index are well-matched for metal-poor
$z \sim 0$ analogs with high electron temperatures. Whatever is causing neon to appear enhanced is
not changing the expected strength of [O~III] $\lambda 5007$.

\subsection{Stacked Measurements}
\label{sec:stack}

\begin{figure}[htp] 
\includegraphics[width=.48\textwidth]{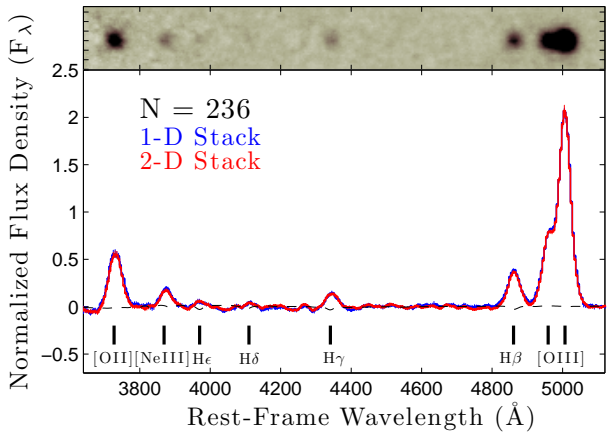}
\centering
\caption{The top panel shows the observed median 2-D spectrum of our entire
$z \sim 2$ galaxy sample, with black indicating higher intensity.  The bottom panel compares the
observed spectrum obtained by summing the central $\pm 3$ pixels of this image (red) to that obtained by median combining all the 1-D extracted spectra (blue).  The error bars, which are
barely visible in this scaling, were calculated using 500 bootstraps with replacement iterations,
computing the median values, and taking the 16th and 84th percentiles of each wavelength bin. 
The spectral resolution of the stacked sample is $\sim$30 \AA.  The location of the emission lines of [O~III] $\lambda 3727$, [Ne~III] $\lambda 3869$, 
[O~III] $\lambda\lambda 4959,5007$, and the Balmer series (H$\beta$, H$\gamma$, H$\delta$, and H$\epsilon$) are marked.  The black dashed curve is the continuum-subtracted stack of the best-fit stellar spectral energy distributions.}
\label{fig:stackcomp}
\end{figure}

In many of our $z \sim 2$ star forming galaxies, [Ne~III] $\lambda 3869$ is too faint
to detect, and by omitting these galaxies from our analysis, we are biasing our sample.  In fact, 
only $\sim 25\%$  of our galaxies have [O~II], [Ne~III], H$\beta$, and [O~III] all
measured with a signal-to-noise ratio greater than one.   Consequently, to understand the rest of the star-forming population, we must rely on spectral stacks of various galaxy subsets, obtained from both the 1-D extracted spectra
and from the original 2-D data.

To perform these stacks, we followed the procedures discussed in \citet{francis+91}.  We began 
by subtracting the continuum from each spectrum using a single fourth-order polynomial in the 1-D case and
multiple fourth-order polynomials covering the spatial direction for the 2-D data.  We then shifted
each spectrum into the rest frame via linear interpolation with 1/3 pixel (5~\AA\ sampling) in the
wavelength direction, and 1/3 pixel ($0\farcs 04$) sampling in the spatial direction.  Slightly
different sampling sizes (i.e., 1/2 or 3/4 of a pixel) did not substantially effect the result.  Next, we
scaled each spectrum to its flux in the [O III] $\lambda$5007 line, and computed the median value for
each wavelength, and (for the 2-D spectral extractions) for each spatial position.  By
median stacking, we preserved the relative strengths of the emission lines, enabling us to
measure line ratios.  We then computed errors for the median stack using the 16th and 84th
percentiles of 500 bootstrap with replacement iterations of this process. Finally, to check for
consistency, we summed the central $\pm 0\farcs 128$ ($\pm 3$ re-sampled pixels, or $\sim 1$~kpc)
of the 2-D median spectrum and compared it to the corresponding median produced from the
1-D analysis. As Figure 4 illustrates, there is excellent agreement between the two results. The
collapsed 2-D median spectrum does have a slightly higher signal-to-noise ratio than the 1-D median,
and we use that in the analysis below. However, the results do not change if we use the 1-D median
stacks.

To investigate the behavior of [Ne~III] $\lambda 3869$ with galaxy type, we divided our
sample of $z \sim 2$ galaxies into subgroups and stacked their spectra by stellar mass, SFR, 
and mass-specific star formation rate (sSFR\null) .   As expected, the results of these analyses are
correlated, since, as can be seen in Figure~\ref{fig:sfrmass}, star formation rate and stellar mass
are related.   Moreover, as Figure~\ref{fig:ssfrmass} illustrates, our
selection criteria produce an artificial correlation between sSFR
and stellar mass at the low
mass end of our distribution.   In what follows, we choose mass as the independent 
parameter, understanding that for our sample, lower mass galaxies will have, on average,
lower SFRs but higher sSFRs.  

\begin{figure*}[htp] 
\includegraphics[width=.75\textwidth]{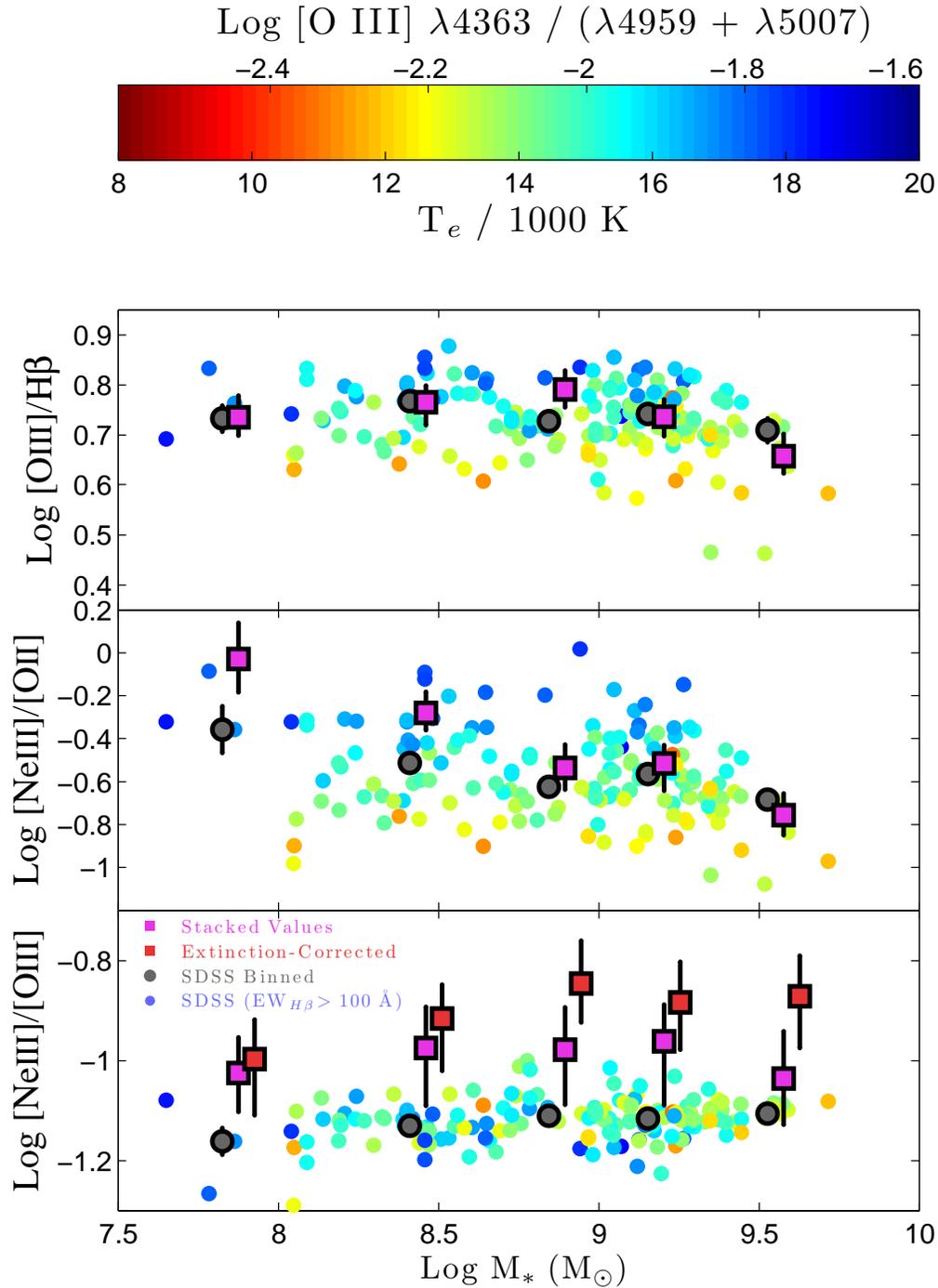}
\centering
\caption{The observed ratios of [O~III]/H$\beta$ (top), [Ne~III]/[O~II] (middle), and
[Ne~III]/[O~III] (bottom) as a function of stellar mass, as derived from SEDs.   The small dots
show the local SDSS comparison galaxies with EW$_{H\beta}$ $>$ 100 \AA, color-coded 
by their [O~III]-based electron temperature, from $T_e = 8,000$~K (red) to 20,000~K (blue). 
The grey circles are the median values for this sample, with their $1 \, \sigma$ uncertainties.
The purple squares denote values obtained from our median stacks of $z \sim 2$ systems, while
the red squares in the [Ne~III]/[O~III] diagram use the galaxies' UV slopes and the
\citet{calzetti01} obscuration law to correct these values for reddening.   The error bars
are computed as in Figure~\ref{fig:ratiomass}, and only include the statistical uncertainties associated
with median stacking.  The [Ne III]/[O III] values in our $z \sim 2$ galaxies are consistently higher
than expected from local samples.}
\label{fig:ratiomasscomp}
\end{figure*}

Table~\ref{tab:stacked} presents the median line ratios for the stacked spectra of five mass bins.
These values are displayed via the purple squares in Figures~\ref{fig:ratiomass}, and 
Figure~\ref{fig:ratiomasscomp}  compares the ratios to measurements of 
[O~III]/H$\beta$, [Ne~III]/[O~II], and [Ne~III]/[O~III] for the well-matched local sample
of high H$\beta$ equivalent width objects.  From the figures, it is clear that the forbidden
neon-to-oxygen line ratios of the $z \sim 2$ universe are offset from those observed in nearby galaxies, 
even when comparing to systems with identical stellar masses and specific star-formation rates.  
In the [Ne~III]/[O~II] diagram, the trend with stellar mass follows that defined by local low-mass
starburst galaxies, but with a mean offset of $\sim 0.12$~dex, which is still marginally consistent with
the high-EW$_{H\beta}$ local sample.  The offset in the [Ne III]/[O III] diagram is more dramatic:
even if one assumes {\it no\/} extinction, there is still a $\gtrsim 0.06$~dex offset between
the $z \sim 2$ and local sample for each mass bin.  If a \citet{calzetti01} obscuration law and the UV slope is used to correct
each grism spectrum prior to stacking, this
offset balloons to $\gtrsim 0.16$~dex across all masses (7.5 $<$ log(M$_*$/M$_{\odot}$) $<$ 10.5).  In contrast, there is no difference between the [O III]/H$\beta$
ratios derived from our median stacks and the local galaxies. Table~\ref{tab:stackedcomp} describes these offsets.

\section{Interpreting the Line Ratios}

Although they are likely oversimplified, photo-ionization models have both qualitatively
and quantitatively reproduced the emission line ratios in the local universe
\citep{kewley+02, dopita+06}.   We can therefore use such models \citep[e.g., CLOUDY
Version 13.03;][]{ferland+98, ferland+13} to gain insight into the dominant physical
parameters which govern emission lines at $z \sim 2$.

To do this, we ran a grid of uniform-density plane-parallel CLOUDY models, fixing the relative
elemental abundances at solar values (in particular, $\log {\rm Ne/O} = -0.76$), while
varying the nebula's oxygen abundance ($-2 < {\rm [O/H]} < 0$), electron density
($10 < n_e < 10^6$~cm$^{-3}$), ionization parameter ($-3.2 < \log U < -1.4$,
where $U$ is the ratio of ionizing photons to the number density of atoms), and
hardness of the ionizing spectrum ($40,000~{\rm K} < T_{\rm eff} < 70,000~{\rm K}$).
We then examined the strengths of the [Ne~III] $\lambda 3869$, [O~II] $\lambda 3727$,
[O~III] $\lambda 5007$, and H$\beta$ lines as a function of these parameters, in order to
explore the possible origins of enhanced neon-to-oxygen ratios.  
 The local analog sample line ratios are best reproduced by CLOUDY models with densities between
10$^2$-10$^3$~cm$^{-3}$, effective black body temperatures of $55,000~{\rm K} < T_{\rm eff} < 65,000~{\rm K}$,
ionization parameters in the range of $-2.8 < \log U < -2.0$, and oxygen abundances between 8.1 $< 12 + \log {\rm O/H} <$8.6.
These values are slightly higher but still consistent with the metallicity range measured via [O~III]$\lambda$4363 and agree
with the best-fit values of the CLOUDY models used by \citet{steidel+14} for their $z \sim 2.3$ sample.      

Our analysis is similar to that performed by \citet{levesque+14}, who mated the MAPPINGS III photo-ionization
code \citep{binette+85, sutherland+93, groves+04} to STARBURST99 \citep{SB99,
vazquez+05} instantaneous burst stellar populations, using the assumption of a
plane-parallel isobaric geometry and $n_e = 100$~cm$^{-3}$.    Their results (as seen
in their Figure~2), overestimate the local relationship between [Ne III]/[O III] by $\sim 0.4$~dex,
and predict line strengths that are even greater than that seen in the $z \sim 2$ universe.
The source of the discrepancy is unknown.

\subsection{Interpretation}

To understand the line ratios displayed in Figures~\ref{fig:ratiomass} and 
\ref{fig:ratiomasscomp}, we begin by considering the local sample of galaxies.  In the diagrams, 
we plot two sets of objects: those whose line ratios suggest the presence of an AGN, and 
those which appear to be normal star-forming galaxies.  The former systems are generally
identified by comparing [O~III]/H$\beta$ to [N~II]/H$\alpha$
\citep[i.e., via the BPT diagram;][]{bpt81}, although in the absence of [N~II]/H$\alpha$, 
stellar mass can be used as a proxy \citep{juneau+11, juneau+14}.  Few of
our objects fall on this AGN sequence; in general, the $z \sim 2$ galaxies detected by
the {\sl HST\/} grism are lower-mass objects with [O~III]/H$\beta$ ratios that place them securely in 
the star-formation region of the diagram.  The star-forming galaxies themselves lie on 
a sequence traced by both the nebular electron temperature and excitation:  as the mass
of a system decreases, the auroral [O~III] $\lambda 4363$ line becomes more prominent,
and the dominant state of oxygen shifts from O$^+$ to O$^{++}$.  Both of these trends 
are generally interpreted as being due to metallicity, as fewer metals mean hotter stars and
larger ionization parameters \citep[e.g.,][]{massey+05},  while fewer coolants translate into higher electron temperatures \citep[e.g.,][]{evans+85, dopita+06}.

The [Ne~III] line ratios derived from the stacked spectra follow these trends.  As the stellar mass 
and metallicity increase, there is a decline in the (reddening independent) ratio of 
[Ne~III]/[O~II] which exactly follows that seen in the local universe.  Meanwhile, the 
(reddening-corrected) [Ne~III]/[O~III] ratio marginally increases by $0.08 \pm 0.05$~dex per 
dex in stellar mass, again, consistent with local observations ($0.02 \pm 0.02$~dex per dex in stellar mass).  

The difficulty arises when considering the absolute strength of [Ne~III] $\lambda 3869$.    Neon and
oxygen are formed together in Type~II supernovae \citep{woosley+95, timmes+95}, and their
doubly-ionized states have similar ionization potentials.  Consequently, in the local universe,
[Ne~III] $\lambda 3869$ and [O~III] $\lambda 5007$ are tightly correlated across all star-forming
galaxies \citep[e.g.,][]{izotov+06, izotov+11}.  The 178 
high-equivalent width (EW$_{{\rm H}\beta} >$ 100 \AA) starburst galaxies which most 
closely match our $z \sim 2$ systems in both stellar mass and specific star formation rate
are plotted in Figure~\ref{fig:ssfrmass}.   
The median de-reddened value of [Ne~III]/[O~III] for these objects is $0.078 \pm 0.001$.  In
contrast, the value for our total median stacked $z \sim 2$ spectrum is $0.091 \pm 0.008$ 
without any correction for reddening.  If we then de-redden each spectrum using the slope of the
stellar UV continuum and a \citet{calzetti01} obscuration law, the 
ratio of the re-stacked data becomes $0.118 \pm 0.011$, i.e., nearly $4 \, \sigma$ greater than the local measurement.  
This offset is higher than that found by \citet{masters+14}, who obtained a $\sim 2\, \sigma$ 
offset of $0.085 \pm 0.003$ using the extinction-corrected composite spectrum of 23 galaxies with 
$8.5 < \log (M/M_{\odot}) < 9.5$ and $\langle z \rangle \sim 1.8$.  \citet{masters+14} did note,
however, that their data have a rather large ($\sim 20\%$) uncertainty in the relative 
calibration of [O~III] to H$\alpha$.   Their measurements are therefore consistent with ours when taking the relative calibration uncertainties
into account.

 The $\sim$1.5$\times$ enhancement in the [Ne~III] emission of $z \sim 2$ star-forming galaxies over that seen in local galaxies with similar global properties, 
has several possible explanations:

\subsubsection{Potential Systematics}  

The high [Ne~III]/[O~III] ratio found for our $z \sim 2$ galaxies
relies on our stacking analysis, so an error in our methodology or extinction correction 
could result in an elevated ratio.  To investigate the former, we scrutinized the flux calibration, continuum subtraction, and biases related to our stacking methodology.  For flux calibration, we relied only on the G141 grism calibration\footnote[7]{www.stsci.edu/hst/wfc3/documents/ISRs/WFC3-2011-05.pdf and www.stsci.edu/hst/wfc3/documents/ISRs/WFC3-2014-01.pdf}, which has been documented to be accurate to $\leq 5\%$.  Nevertheless, to  test the relative flux calibration over scales of $\sim$2000 \AA, we binned our data by redshift (or, equivalently, by observed wavelength) and found no significant variations in the ratios of emission line strengths, suggesting that the flux calibration is indeed stable.  Similarly, to test for errors in our continuum subtraction, we binned our data by estimated levels of spectral contamination.  In our 1-D and 2-D grism spectra, the underlying continuum is a combination of stellar emission from the galaxy and residual contamination from overlapping spectra (see Section~\ref{sec:sample}).  To capture
the trends in both, we fit the continuum with a fourth order polynomial, and this could, in theory, introduce a systematic bias. However, an analysis of our spectra shows no significant relation between line strength and contamination.  Moreover,  
when we repeated our stacking procedure using the best-fit SEDs from our stellar mass estimations, the simplified polynomial approximation enhanced [Ne~III] (by 5\%), leaving [O~II] and [O~III] relatively unaffected.

To verify the robustness of our stacking procedure, we compared the line ratios found from
the median stacks of all the objects with individual line detections with the median of the
individual line ratios measured from these same data.   The results were consistent.
Also, to examine whether our extinction corrections were affecting the stacking process, we
compared the results of stacks where reddening corrections were applied prior to stacking
with similar stacks where de-reddening occurred only after the fact. Again, there was no
difference in the results.  Finally, one complication related to WFC3 IR grism extractions
is that the point spread function (PSF) becomes slightly larger with wavelength. For point
sources this must be taken into account, but if the galaxies are substantially larger than
the PSF, the issue is mitigated.  As seen in the cutouts of Figure 1, most of our $z \sim 2$
galaxies are indeed resolved.  An examination of the spatial dependence of our
2-D stacked spectra confirms that the extraction algorithm is not introducing a significant
systematic error into our line measurements:  the spatial extent of [O~II] and [O~III], the two
emission lines with the largest wavelength separation, are consistent 
($\sim  0\farcs 49$ full-width half maximum for the total stack).

The significance of our [Ne~III]/[O~III] result also depends on how we correct our
spectra for the effects of dust.  Due to the excellent photometric coverage in the rest-frame
UV, the uncertainties associated with our stellar reddenings are relatively small, and
this term contributes only $\sim 30\%$ of the total [Ne~III]/[O~III] error budget.
However, the possible systematic uncertainty associated with our nebular extinctions
could be larger.  To translate our stellar reddenings into nebular extinctions,
we used the \citet{calzetti01} obscuration law, which states that the applicable extinction
for nebular lines is greater than that for starlight by $E(B-V)_{\rm stars} = 0.44 E(B-V)_{\rm gas}$.
This result is supported by numerous comparisons of UV- and Balmer line-based SFR
measurements in the $z \gtrsim 2$ universe \citep{forster+09, mannucci+09, wuyts+13,
price+13, holden+14, zeimann+14}, but counter-examples exist as well \citep{erb+06}.  Thus, we can take the very
conservative approach and assume that the stars and gas are equally extinguished.  Under this
prescription, the ratio of [Ne~III]/[O~III], $0.107 \pm 0.010$, is still discrepant with the local
value at the $\sim 3\,\sigma$ level.  It is therefore exceedingly unlikely that the high
neon ratios seen at $z \sim 2$ are due to our assumptions about reddening.

\subsubsection{Collisional De-Excitation}

\begin{figure*}[htp] 
\includegraphics[width=.90\textwidth]{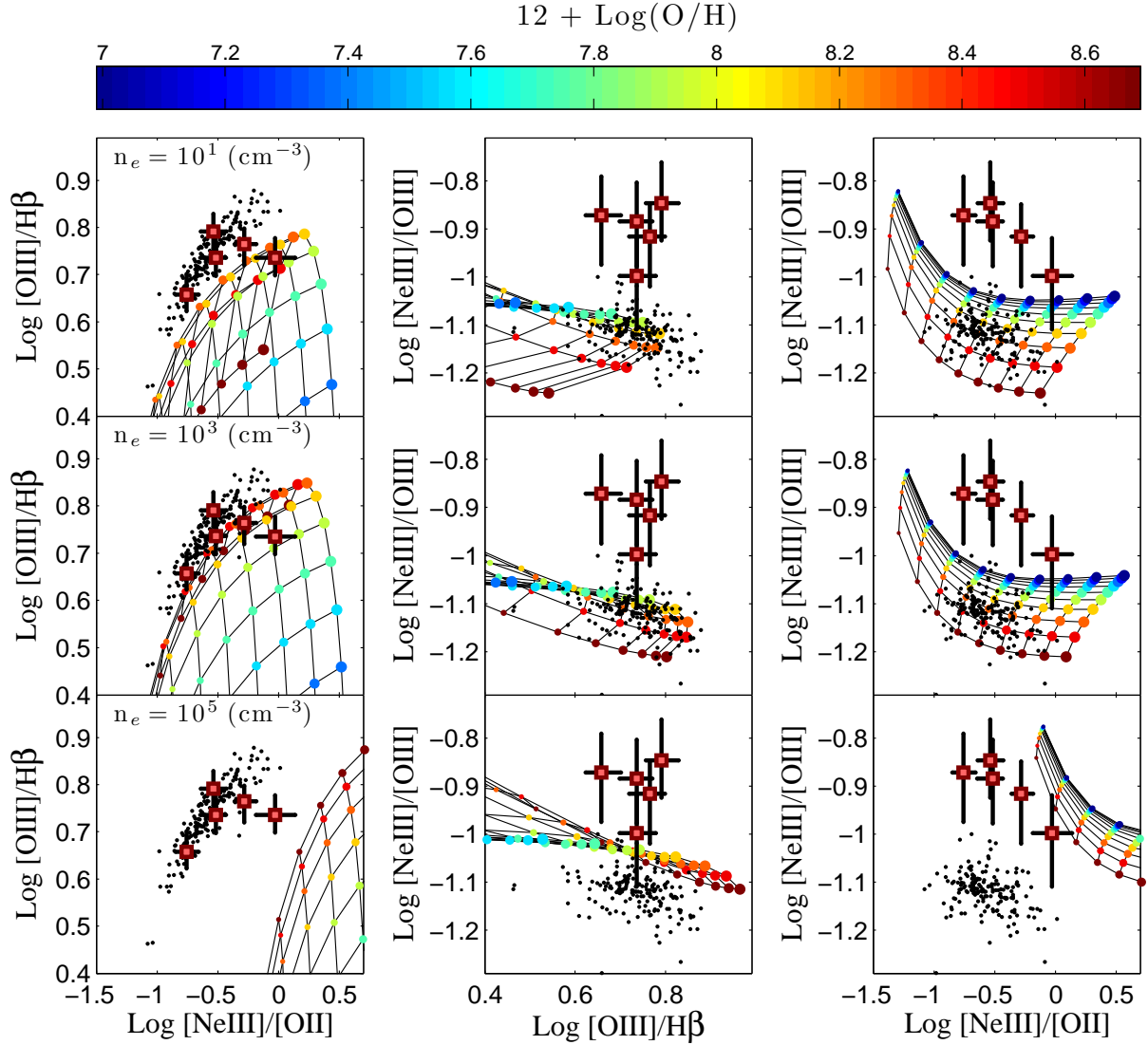}
\centering
\caption{The line ratios of our high H$\beta$ equivalent width local galaxies (black dots) and 
our stacks of $z \sim 2$ emission-line galaxies (red squares) compared to the
predictions of CLOUDY models in three different projections of our emission-line space. 
[Ne~III]/[O~III] has been corrected for extinction using the \citet{calzetti01} obscuration law.  
The points on the CLOUDY grid are color-coded by metallicity following the scale at the top of 
the diagram, with the  size of each point proportional to the value of the ionization parameter.  
The temperature of the ionizing source has been held fixed at 55,000~K.   From top to bottom 
the grids represent densities of 10, $10^3$, and $10^5$~cm$^{-3}$.   Although the lower 
density models can reproduce the line ratios of the local galaxies reasonably well, no model
can fit the $z \sim 2$ data.}
\label{fig:ratiocloudydens}
\end{figure*}

 The critical density of the 2$P$ $^1D_2$ state of O$^{++}$ that
produces [O~III] $\lambda 5007$ is $\sim 10^6$~cm$^{-3}$; for the equivalent state of 
Ne$^{++}$, this density is $\sim 10^7$~cm$^{-3}$.  Hence if the density of a typical $z \sim 2$ 
H~II  region is significantly greater than that observed for H~II regions in the local universe 
(i.e., $n_e \gtrsim 10^5$~cm$^{-3}$), collisional de-excitation of oxygen will cause [Ne~III]
$\lambda 3869$ to appear enhanced.   Such densities are more extreme than what is usually seen
in massive $z \sim 3$ galaxies \citep{shirazi+14}, but are still plausible, especially if the star-forming
regions follow a size-density relation \citep[e.g.,][]{kennicutt84, hunt+09}.   Moreover, high 
densities have been invoked to explain the strengths of mid-infrared emisson lines in the 
Antennae galaxies \citep{snijders+07}, as well as the emission from the molecular gas of 
ultra-luminous infrared galaxies \citep{solomon+92}.  The effect of electron density on our line 
ratios can be seen in Figure~\ref{fig:ratiocloudydens}.  We display, from left to right, the three projections of our three dimensional  emission-line space and overlay our CLOUDY models 
for three different densities: 10, 10$^3$, 10$^5$ cm$^{-3}$.   For this analysis, the ionizing
spectrum has been fixed to a 55,000 K blackbody spectrum as suggested in \citet{steidel+14}. 
The figure demonstrates that densities of $\lesssim 10^3$~cm$^{-3}$ can successfully explain 
the line ratios of local galaxies, with perhaps a small deficit of [O~III].  However, no value of abundance, ionization parameter, or density can simultaneously explain all three line ratios 
for the $z \sim 2$ systems.

The difficulty with invoking high densities is that the collisional de-excitation of oxygen should
affect other lines as well.  For example, if [O~III] emission were decreased by the requisite
$\sim 50\%$, the [O~III]/H$\beta$ ratio would be lowered by that same factor.   This decrease is 
not evident in the top left panel of Figure~\ref{fig:ratiomass}.  Similarly, because the critical
density for the nebular lines of [O~II] is $\sim 10^3$~cm$^{-3}$,  the [Ne~III]/[O~II] ratio would be
{\it greatly\/} enhanced in a high-density environment (by at least a factor of $\sim 5$; see Figure~\ref{fig:ratiocloudydens}).   
While it may be possible to design a multi-zone model 
wherein [O~II] and [O~III] are produced in different regions of a galaxy, substantial tuning would be
needed, and the [O~III]/H$\beta$ discrepancy would remain.

\subsubsection{Hotter Exciting Stars}  

\begin{figure*}[htp] 
\includegraphics[width=.90\textwidth]{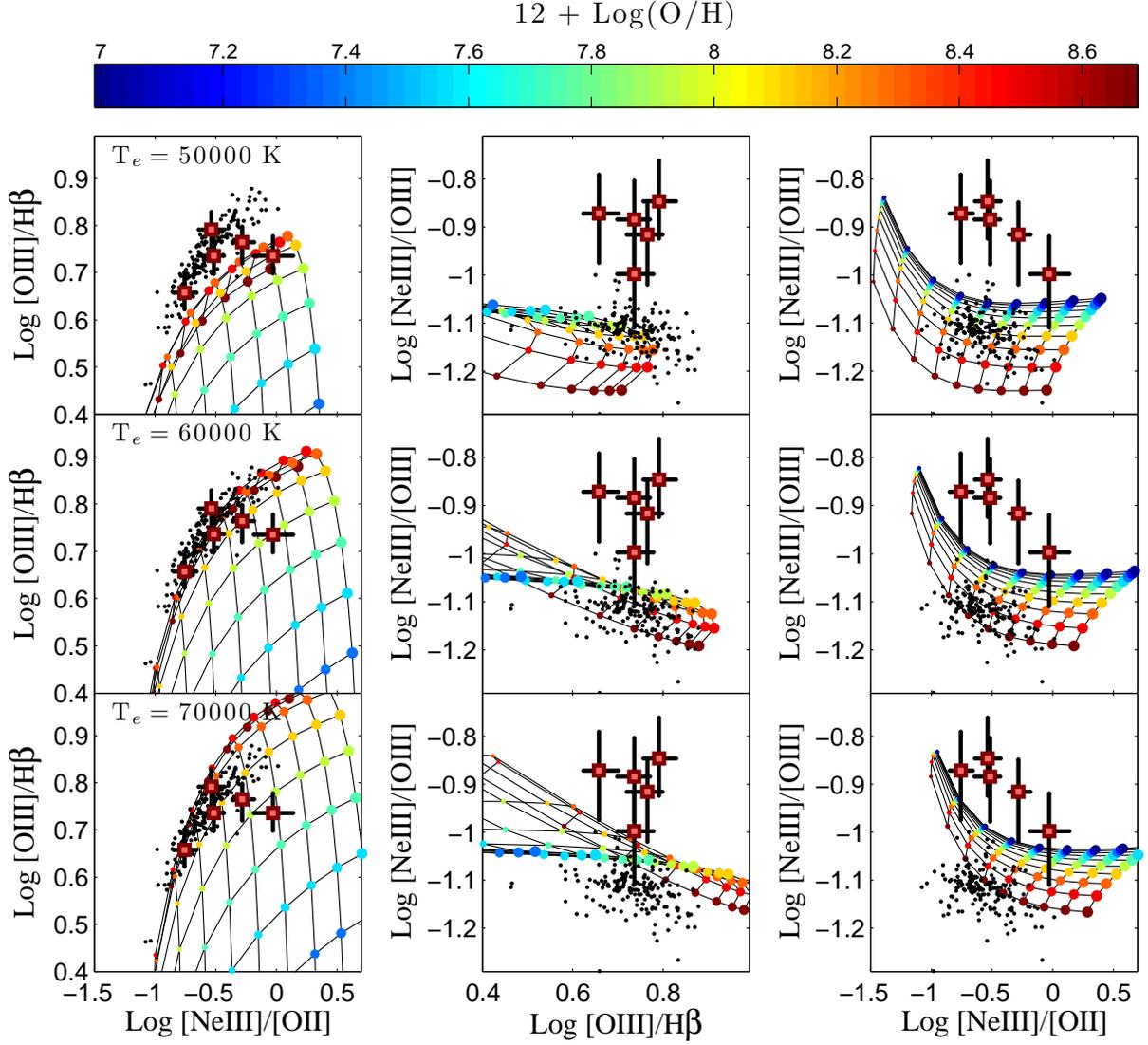}
\centering
\caption{The line ratios of our high H$\beta$ equivalent width local galaxies (black dots) and 
our stacks of $z \sim 2$ emission-line galaxies (red squares) compared to the
predictions of CLOUDY models in three different projections of our emission-line space. 
The [Ne~III]/[O~III] measurements have been corrected for extinction using the \citet{calzetti01} obscuration law.  
The points on the CLOUDY grid are color-coded by metallicity following the scale at the top of 
the diagram, with the  size of each point proportional to the value of the ionization parameter.  
The density of the ISM has been held fixed at $10^3$~cm$^{-3}$.   From top to bottom 
the grids represent effective temperatures of 50,000, 60,000, and 70,000~K.   Despite increasing the blackbody temperature to 70,000 K, the models are still unable to reproduce 
our stacked values.}
\label{fig:ratiocloudytemp}
\end{figure*}

For more than a decade, it has been clear that high-redshift galaxies 
have, on average, higher [O~III]/H$\beta$ ratios than similar galaxies in the local universe
\citep[e.g.,][]{teplitz+00}.  With the largest sample of $z \sim 2.3$ galaxies observed to date,
\citet{steidel+14} showed that, at any given [N~II]/H$\alpha$, higher than normal 
[O~III]/H$\beta$ ratios could be obtained by increasing the temperature of the ionizing source
from the assumed value of $\sim 42,000$~K for the local SDSS galaxies, to $\sim 55,000$~K.   
Since Ne$^{++}$ has a higher ionization potential than O$^{++}$, this change could, in theory, increase 
the fraction of doubly-ionized neon relative to oxygen.  However, the difference in the ionization
potentials of the two species is only 5.9~eV (41.0~eV versus 35.1~eV), so once beyond 
45,000~K, increasing the temperature of a blackbody source has little effect on the ratio of the
ionization states.  This result is illustrated in Figure~\ref{fig:ratiocloudytemp}.

Galaxies, of course, are not a single blackbody sources, but a composite of many different
stars with different effective temperatures.  To examine this effect, we used the STARBURST99 (SB99, version 6.0.4) population synthesis code \citep{SB99, vazquez+05} to create a more
realistic ionizing spectrum for CLOUDY input.  For a constant star formation history, the ionizing spectrum of a SB99 galaxy is insensitive to age, once the system is older than $\sim 10^7$~yr.
Indeed, if all the other parameters are fixed, the ionizing spectrum is well-fit by a blackbody
with an effective temperature primarily determined by the stellar metallicity.  For a 
$Z = 0.02 Z_{\odot}$ system, this temperature is roughly 65,000~K; for $Z = 0.2 Z_{\odot}$, 
it is 55,000~K, and varying the high-mass slope of the IMF has little effect on this result.  
We did not explore binary population synthesis or the effects of stellar rotation, but it is clear that standard SB99
models cannot produce a spectrum that is hard enough to match the observed line ratios.

\subsubsection{Oxygen Depletion Onto Dust Grains}  

While neon is an inert gas, oxygen is reactive, and can be depleted out of the gas phase
onto dust grains or molecules. In the local universe,  $\sim 40\%$ of oxygen is thought to be
locked up in grains and molecules such as water and O$_2$ \citep{jenkins09}; if this ratio is higher in
the $z \sim 2$ universe, then our [Ne III] $\lambda$3869 observations can be explained.  However, one
expects fewer opportunities for dust depletion at $z \gtrsim 2$, not more.
Conventionally, oxygen is bound up by refractory compounds such as metallic oxides and
amorphous silicates, which require the prior existence of $\alpha$-process and iron-peak elements.
In the metal-poor environment of our $z \sim 2$ systems, this process must be inhibited.  On the
other hand, when oxygen consumption is measured against the abundance of metals such as
Mg, Si, and Fe, its depletion from the gas phase is six times higher than expected (Jenkins 2009).
This suggests the existence of an additional avenue for the removal of oxygen from the gas phase.
However,  since this second avenue probably involves the creation of ices on the surfaces of dust
grains \citep{ioppolo08}, it is still likely to depend on metallicity.   Still, the possibility of dust
depletion cannot be ruled out.

\subsubsection{Enhanced Neon to Oxygen Abundance}  

The most straightforward interpretation is that the neon to oxygen abundance ratio in
the $z \sim 2$ universe is enhanced by $ \sim 0.2$~dex over that of the local universe.
At face value, this scenario seems unlikely.  In the Sun, the dominant isotope of neon
is that of $^{20}$Ne, which contributes 92.9\% of all neon atoms, with $^{22}$Ne being
the next most abundant species \citep[6.8\%;][]{asplund+09}.  Since $^{20}$Ne
is formed directly from $^{16}$O via $\alpha$-capture, the formation regions
of the two species must be highly correlated.  Indeed, abundance measurements in Milky
Way and Magellanic Cloud planetary nebulae \citep{henry89}, H~II regions \citep{garnett02},
and other galaxies \citep{izotov+06, izotov+11} yield the same cosmic Ne/O abundance
over a wide range of metallicity ($7.1 < 12 + \log {\rm O/H} < 8.4$).

On the other hand, $^{22}$Ne has a different formation site:  the hydrogen and helium
burning shells of massive stars in their late stages of evolution.  During a thermal pulse, some of
the $^{12}$C and $^{16}$O produced by the fusion of helium can be convected outward to the
site of the temporarily dormant hydrogen burning shell.   When the hydrogen re-ignites, this
material will be converted to $^{14}$N via CNO burning, and, upon the next thermal pulse, some of
this nitrogen will be transported back to into the helium burning shell, where it will undergo
$\alpha$ captures.  The result is the creation of $^{18}O$, $^{22}$Ne, $^{25}$Mg, and s-process
elements.  Dredge-up can then bring this material to the stellar surface, where it can be lost to the
interstellar medium via stellar winds.  If the importance of this  process is greater in the $z \sim 2$
universe, the result can be an enhancement in the total neon to oxygen ratio.

Interestingly, the analysis of Galactic cosmic rays suggests a $^{22}$Ne/$^{20}$Ne ratio of
$\sim 0.4$, i.e., a value that is $\sim 5.3$ times higher than what is found in the solar wind
\citep[e.g.,][]{binns05}.  Although the cause of this offset is still debated, the most widely
accepted explanation for the excess $^{22}$Ne ions lies in the physics of carbon-rich
Wolf-Rayet stars  \citep{casse82, prantzos87}.  The high-velocity winds of these systems cannot 
only deposit the $^{22}$Ne from shell mixing into the local ISM, but the wind interactions with
standing shocks from nearby Wolf-Rayet and precursor OB stars can accelerate material
to the high speeds associated with cosmic rays.  Wolf-Rayet star winds have also been invoked
to explain the high nitrogen to oxygen ratios observed at $z \sim 2$ \citep[e.g.,][]{masters+14,
shapley14}, though in this case, it is the nitrogen-rich WN stars which are the source of the
enhancement.

If, indeed, $^{22}$Ne is solely responsible for the elevated [Ne~III]/[O~III] line ratios observed in
our {\sl HST\/} grism galaxies, then the $^{22}$Ne/$^{20}$Ne isotopic ratio at $z \sim 2$
must $\sim 0.6$.   This is roughly ten times larger than that ratio seen in the Solar wind,
and $\sim 1.5$ times larger than that measured for cosmic rays.  At face value, such an
enhancement does not seem plausible.  However, many aspects of massive star stellar
evolution, such as the evolution of the IMF with redshift, the effects of stellar rotation, and
the importance of binary stars at high-$z$, are not well known.  Moreover, it is not clear
whether the solar ratio of $^{22}$Ne/$^{20}$Ne is representative of the $z \sim 0$ universe
as a whole:  it is quite possible that the isotopic ratio in spiral arms is larger than that observed
in the solar system.  Thus, this possibility cannot be ruled out.

\subsubsection{Compton-Thick AGN}  

Non-thermal radiation can enhance [Ne~III] relative
to [O~III] and reproduce the line ratios seen at $z \sim 2$.   As the bottom left panel of
Figure~\ref{fig:ratiomass} demonstrates, the local sample of galaxies classified as 
AGN by the MEx diagnostic \citep{juneau+11, juneau+14} clearly have enhanced values of
[Ne~III] relative to the locus of star-forming galaxies.  

All three of our program fields --- GOODS-N, GOODS-S, and COSMOS -- have medium deep
and deep X-ray data \citep{elvis+09, alexander+03, xue+11}, and, at the redshifts considered 
here ($1.90 < z < 2.35$), X-rays associated with normal star formation will be well 
below the limits of detection \citep{lehmer+10}.  Consequently, any emission-line galaxy whose position is co-incident with an X-ray source is likely powered by an AGN.  

To search for these AGN, we cross-correlated our $z \sim 2$ object catalog with the list of 
X-ray sources found in the COSMOS, GOODS-N, and GOODS-S regions.   Only four of our 
emission-line  galaxies lie within $2.5\arcsec$ of an X-ray source, and, in all four, the 
signal-to-noise ratio is too low to place a significant limit on [Ne~III]/[O~III]\null.  We then stacked the
remaining 63 galaxies in the Chandra Deep Field South 4Ms field \citep{xue+11} following the
procedure described by \citet{luo+11}, and examined the resulting image for evidence of X-ray
emission.  No X-rays were found down to 90\% upper limits of $3.2 \times 10^{-18}$, 
$6.7 \times 10^{-18}$, and $6.8 \times 10^{-18}$~ergs~s$^{-1}$~cm$^{-2}$ for the 0.5-2.0 keV, 
2.0-8.0 keV, and 0.5-8.0 keV bandpasses, respectively.  These limits correspond to a maximum
absorbed X-ray luminosity of $\sim 2 \times 10^{41}$~ergs~s$^{-1}$ at the average redshift 
of the stacked sample ($z = 2.09$) and constrain
the maximal fractional contribution of AGN to the observed [O~III] and [Ne~III] luminosities.

\begin{figure}[htp] 
\includegraphics[width=.48\textwidth]{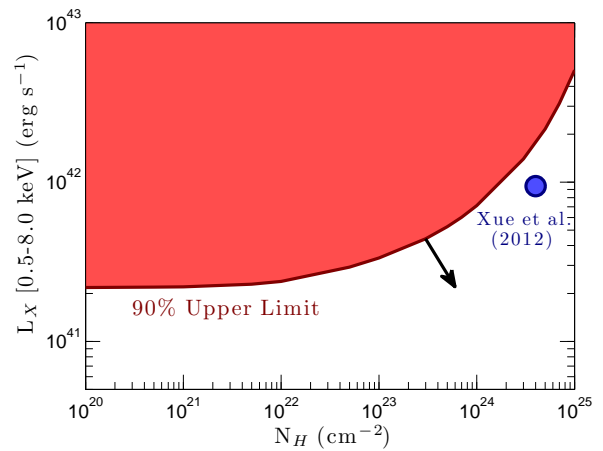}
\centering
\caption{The intrinsic X-ray luminosity for 0.5-8.0 keV (L$_X$) is plotted against the absorbing hydrogen column density (N$_H$) for the 90\% upper limit found in our X-ray stacking analysis.  The upper limit curve is plotted in dark red and the region ruled out by the stacking analysis is shown in light red.  The arrow indicates the phase-space still allowed.  \citet{xue+12} performed an X-ray stacking analysis and found that Compton-thick AGN (N$_H$ $\sim$ 4$\times$10$^{24}$ cm$^{-2}$) at $z = 1 - 3$ with $2\times10^8 < $M$_*$/M$_{\odot}$ $< 2\times10^9$ likely produce 20\%-25\% of the 6-8 keV cosmic X-ray background.  The stacked X-ray luminosity of these sources is shown as a blue dot.}
\label{fig:lxnh}
\end{figure}

\citet{xue+12} has argued that 20-25\% of the cosmic hard X-ray background (6-8 keV) emanates 
from heavily obscured AGN residing in low mass ($2 \times 10^8$ to $2 \times 10^9 \, M_{\odot}$), high redshift ($1 < z < 3$) galaxies, with an average 0.5-2 keV flux of
$1.43 \times 10^{-18}$~erg~cm$^{-2}$ s$^{-1}$.  This is below the 90\% upper limit found in our stacking analysis, so it remains possible that the high [Ne~III]/[O~III] ratios are being caused by obscured AGN.  This is illustrated in Figure~\ref{fig:lxnh}, which uses the {\sl Chandra\/}
Proposal Planning Toolkit\footnote[8]{http://cxc.harvard.edu/toolkit/pimms.jsp} to show our intrinsic X-ray luminosity limit as a function
of absorbing column density, $N_H$, for an AGN with photon power law index of $\Gamma = 1.9$.

To better quantify the effect of AGN on the observed [Ne~III]/[O~III] ratio, we can begin by considering
our median $z \sim 2$ galaxy, which has an extinction-corrected  [O~III] luminosity of
$\sim 5 \times 10^{42}$~ergs~s$^{-1}$.  Compton-thin AGN with 2-8~keV luminosities near the
limit of our stacking analysis ($L \sim 2 \times 10^{41}$~ergs~s$^{-1}$) typically have [O~III] to
X-ray luminosity ratios of between $\sim 0.01$  and 0.1 \citep{lamastra+09, wilkes+13}.  This means that
these objects can contribute no more $\sim 0.4\%$ to the total [O~III] luminosity of our stacked galaxy.
In contrast, Compton-thick AGN can have [O~III] to X-ray luminosity ratios that are $\sim 100$ times larger
\citep{wilkes+13}, hence their contribution could conceivably be as large as $\sim 40\%$.  If we let
$\eta_{\rm SF} = 0.078 \pm 0.001$ be the mean [Ne~III]/[O~III] ratio for our local sample of
high-equivalent width star forming galaxies, and $\eta_{\rm AGN} = 0.21$ be the same quantity for
local Seyfert~1 AGN \citep{nagao+01}, then the measured [Ne~III]/[O~III] ratio in our $z \sim 2$
stacked galaxy should be
\begin{equation}
\eta_{\rm obs} = \frac{\eta_{\rm SF} + f \eta_{\rm AGN}}{1+f}
\end{equation}
where $f$ is the median fractional contribution of AGN to the luminosity of [O~III]\null.  For $f \lesssim 0.4$,
$\eta_{\rm obs} \lesssim 0.116$, so this is still consistent with the measured values of our
stacks.  This argument holds for any distribution of the fractional contribution of AGN to the 
luminosity of [O~III] as long as the median of that distribution is a few tens of percent.  
We therefore cannot rule out the hypothesis that an unseen population of Compton-thick AGN
is causing the $z \sim 2$ [Ne~III]/[O~III] line ratios to appear enhanced.

\section{Future Avenues}
When comparing the median stacked values for our 236 $z \sim 2$ galaxies to a local 
comparison sample from SDSS (well-matched in stellar mass and sSFR), we find consistent 
[O~III]/H$\beta$ ratios across all stellar masses.  However, [Ne~III]/[O~II] is elevated by
$0.23 \pm 0.13$~dex for $\log M_*/M_{\odot} \lesssim 9$ and [Ne~III]/[O~III] is persistently
high by more than $1.5 \, \sigma$ (0.16 to 0.25~dex) for each mass bin and all reasonable 
extinction laws with a total significance of $3.6 \, \sigma$.  There are several possible mechanisms that can explain the high ratios, 
including oxygen depletion, extreme densities, and the existence of a large population of
X-ray obscured AGN.   Further observations with current ground-based 8-m class telescopes as well as upcoming space-based missions like the {\it James Webb Space Telescope} are required to assess whether the elevation is 
truly compelling and to identify the physical reasons for it. 

The ratio of [Ne~III]/[O~III] is sensitive to dust, and although we were 
able to estimate the attenuation through the indirect method of stellar reddening, measurements of the Balmer
decrement would allow an accurate correction.  H$\beta$, H$\gamma$, and
H$\delta$ are all observable in our stacked spectra, but due to their low signal-to-noise ratio,
short lever arm, and the effects of stellar absorption, these ratios yield highly uncertain extinction estimates.  
A measurement of H$\alpha$ placed on the same relative flux scale as H$\beta$ can
be obtained from the ground via the MOSFIRE instrument on Keck \citep{mclean+12} and
the FLAMINGOS-2 spectrograph on Gemini \citep{eikenberry+12}.  This would remove the
uncertainty associated with extinction.

We explored the possibility that the ratio of [Ne~III]/[O~III] could be caused by collisional de-excitation of [O~III] at high
electron densities ($\gtrsim 10^5$~cm$^{-3}$).  This would only be possible in a multi-zonal model as a single zone ionized region cannot explain all three observed line ratios.  If indeed H~II regions in the $z \sim 2$ universe are clumpier than their local counterparts, high critical density UV emission lines such as CIII]$\lambda$1909 may show strong emission.  A CIII] equivalent width of 10\AA\ \citep{stark+14} would translate into a median expected flux of $\sim$3$\times 10^{-18}$~ergs~s$^{-1}$~cm$^{-2}$. 


Further investigations of the [Ne~III]/[O~III] ratio are also possible for
$1.3 \lesssim z \lesssim 1.6$ using GOODS-N archival data from the WFC3 G102
(PI Barro: GO-13420) and G141 grisms.  The G102 grism also provides a measurement of
the higher ionization line [Ne~V] $\lambda 3426$ for nearly 100 of the galaxies in our
sample.  Since AGN exhibit [Ne~V]/[Ne~III] ratios that are an order of magnitude
greater than that of normal star-forming systems, these data could be used to test
whether obscured AGN are responsible for enhanced [Ne~III] emission in the $z \sim 2$ universe.    

\acknowledgements
We would like to thank the anonymous referee for their fantastic input and comments on the manuscript.
This work was supported via NSF through grant AST 09-26641 and AST 08-07873.  The Institute for Gravitation and 
the Cosmos is supported by the Eberly College of Science and the Office of the Senior Vice President for 
Research at the Pennsylvania State University.  This work is based on observations taken by the 3D-HST Treasury Program (GO 12177 and 12328) with the NASA/ESA HST, which is operated by the Association of Universities for Research in Astronomy, Inc., under NASA contract NAS5-26555.

\clearpage

\begin{deluxetable*}{ccccccccc}
\centering
\tabletypesize{\tiny}
\tablecaption{Physical Characteristics of Median Stacked Spectra}
\tablewidth{0pt}
\tablehead{
 \colhead{$\langle$Log M$_{*}$$\rangle$} & \colhead{Num.} & \colhead{$\langle$SFR$\rangle$} &  \colhead{$\langle$Log sSFR$\rangle$} & \colhead{$\langle$$\beta$$\rangle$} &  \colhead{Log [O~III]/H$\beta$} & \colhead{Log [Ne~III]/[O~II]} & \colhead{Log [Ne~III]/[O~III]} & \colhead{Log [Ne~III]/[O~III]}\\
\colhead{M$_{\odot}$} & \colhead{---} & \colhead{M$_{\odot}$ yr$^{-1}$} &  \colhead{yr$^{-1}$} & \colhead{---} & \colhead{---} &\colhead{---} & \colhead{Obs.} & \colhead{Cor.\footnotemark[1] }
 }
 \startdata
7.87 & 47 & 2.0 & $-$7.47 & $-$2.04 & 0.74$\pm$0.05 & $-$0.03$\pm$0.16 & $-$1.02$\pm$0.09 & $-$1.00$\pm$0.10 \\
8.45 & 47 & 3.7 & $-$7.93 & $-$1.98 & 0.76$\pm$0.04 & $-$0.28$\pm$0.09 & $-$0.97$\pm$0.10 & $-$0.92$\pm$0.09\\
8.89 & 48 & 8.2 & $-$8.01 & $-$1.74 & 0.79$\pm$0.04 & $-$0.54$\pm$0.11 & $-$0.98$\pm$0.10 & $-$0.85$\pm$0.08\\
9.20 & 47 & 11.8 & $-$8.08 & $-$1.80 & 0.74$\pm$0.04 & $-$0.52$\pm$0.11 & $-$0.96$\pm$0.10 & $-$0.89$\pm$0.09\\
9.58 & 48 & 32.6 & $-$8.07 & $-$1.53 & 0.66$\pm$0.05 & $-$0.76$\pm$0.10 & $-$1.04$\pm$0.09 & $-$0.87$\pm$0.09\\
\enddata
\tablenotetext{1}{ [Ne~III]/[O~III] corrected for extinction via a \citet{calzetti01} obscuration law using the UV slope.}
\label{tab:stacked}
\end{deluxetable*}

\begin{deluxetable*}{cccccccc}
\centering
\tabletypesize{\scriptsize}
\tablecaption{Line Ratio Comparison between $z \sim 2$ and $z \sim 0$}
\tablewidth{0pt}
\tablehead{
 \colhead{Mass Bin} & \colhead{Num.} & \colhead{Num.} & \colhead{Log [O~III]/H$\beta$}  & \colhead{Log [Ne~III]/[O~II]} & \colhead{Log [Ne~III]/[O~III]\footnotemark[2]}\\
\colhead{Log (M$_{*}$/M$_{\odot}$)} & \colhead{$z \sim 2$} & \colhead{$z \sim 0$} & \colhead{Offset\footnotemark[1] } & \colhead{Offset\footnotemark[1] } & \colhead{Offset\footnotemark[1]}
 }
 \startdata
7.50 - 8.18  & 47 & 11 & $+$0.00$\pm$0.05 & $+$0.33$\pm$0.20 & $+$0.16$\pm$0.10  \\
8.18 - 8.72 & 47  & 43 & $-$0.00$\pm$0.04 & $+$0.23$\pm$0.10 & $+$0.22$\pm$0.09 \\
8.72 - 9.07 & 48  & 40 & $+$0.06$\pm$0.04 & $+$0.09$\pm$0.11 & $+$0.26$\pm$0.08\\
9.07 - 9.35 & 47  & 59 & $-$0.01$\pm$0.04 & $+$0.05$\pm$0.11 & $+$0.23$\pm$0.09 \\
9.35 - 10.50 & 48 & 25  & $-$0.05$\pm$0.05 & $-$0.07$\pm$0.11  & $+$0.23$\pm$0.09 \\
\enddata
\tablenotetext{1}{The difference between the logarithm of the median line ratios for the 
$z \sim 2$ sample and the local comparison sample with high equivalent widths.}
\tablenotetext{2}{ [Ne~III]/[O~III] corrected for extinction via a \citet{calzetti01} obscuration law.}
\label{tab:stackedcomp}
\end{deluxetable*}

\end{document}